\documentstyle[fleqn,epsf,11pt]{article}

\newcommand{\AmS}{{\protect\the\textfont2
  A\kern-.1667em\lower.5ex\hbox{M}\kern-.125emS}}

\begin{document}
\title{The Phase Diagram of Crystalline Surfaces}

\author{K.N.~Anagnostopoulos$^{\rm a}$,\\ 
      M.J.~Bowick$^{\rm b}$,\\ 
      S.M.~Catterall$^{\rm b}$,\\ 
        M.~Falcioni$^{\rm b}$\footnote{Poster 
presented by M.~Falcioni at Lattice '95}\\
        and \\
        G.~Thorleifsson$^{\rm b}$\footnote{
         {\tt bowick},
         {\tt smc}, 
         {\tt falcioni}, 
         {\tt thorleif@npac.syr.edu}, {\tt
konstant@nbi.dk}}\\[2em]
{\small 
a) Niels Bohr Institute, Blegdamsvej 17,}\\
{\small DK-2100 K\o benhavn \O, Denmark.}\\
{\small b) Department of Physics, Syracuse University,}\\
{\small Syracuse, NY 13244-1130
U.S.A.}
}

\maketitle

\begin{abstract}
We report the status of a high-statistics Monte Carlo simulation of
non-self-avoiding crystalline surfaces with extrinsic curvature on
lattices of size up to $128^2$ nodes.  We impose free
boundary conditions.  The free energy is a gaussian spring tethering
potential together with a normal-normal bending energy.
Particular emphasis is given to the behavior of the model in the
cold phase where we measure the decay of the normal-normal correlation
function.
\end{abstract}
\vskip 4em
\begin{flushright}
  {SU-HEP-95-4241-620}\\
  {UFIFT-HEP-95-18}
\end{flushright}
    
\newpage

\section{Introduction}
In recent years there has been a lot of interest in the statistical
mechanics of crystalline and fluid surfaces \cite{Jerusalem1}. The
former is believed to describe physical polymerized membranes
\cite{Schmidt} and the latter may be a regularization of string theory. 

We focus our study on crystalline surfaces with bending rigidity
embedded in $R^3$. It is conjectured that this model has a second order
phase transition driven by the competition between entropy and the
bending energy \cite{NP}. The high temperature phase is characterized
by crumpled configurations. In the low temperature phase the system is
no longer isotropic and the surfaces are roughly flat. 

One may wonder what stabilizes the flat phase. The theoretical
argument is that the in-plane elastic constants prevent the surface
from fluctuating arbitrarily in the embedding space.  
This leads to an effective long wave stiffening of the surface
\cite{Jerusalem1,NP}.

The crumpling transition has been studied numerically with
simulations explicitly incorporating 2-d elastic constants
\cite{KN,KKN,AN}.  There have also been simulations with a simple
gaussian spring potential playing the role of the tethering potential
\cite{ADJ,BEW,RK,HW,WS,BET}.  In both cases evidence has been presented for
a continuous phase transition. 

In the latter class of models the equilibrium spring length is taken
to be zero, and a simple calculation indicates that the microscopic
elastic constants vanish \cite{N}. It is tempting to argue, therefore,
that the flat phase of these models is not truly stable, even in the
limit of large bending rigidity. 

Our ultimate aim is to carefully compare the behavior of the
appropriate observables in the cold phase as a function of the
equilibrium spring length.  

\section{The Model}
Consider a system of particles connected to form a triangular 2--d
mesh embedded in 3 dimensions. Let each particle be labeled by an
internal discrete coordinate system ${\bf x} = (x_1,x_2)$ denoting its
position on the mesh.  Its actual position in the embedding space is
given by the 3 dimensional vector ${\bf r}(x_1,x_2)$.  The action has
a tethering potential and a bending energy term.  Our choice is to use 
simple gaussian springs between the vertices as a tethering potential
and a normal-normal interaction as the bending energy term.  Therefore the
action is

\begin{equation}
S = \sum_{\langle {\rm x x'} \rangle} \left( l_{\rm x x'} \right)^2 +
\lambda \sum_{\langle \alpha\beta \rangle} \left( 1 - \vec{n}_{\alpha}
\cdot \vec{n}_{\beta} \right).
\label{action}
\end{equation}
Here the subscripts ${\rm x, x'}$ label the vertices and $l_{\rm x
x'}$ is the distance between the vertices ${\rm x}$ and ${\rm x'}$ in the
embedding space.  The subscripts $\alpha,\beta$ label the faces
(triangles) of the surface, $\vec{n}$ is the unit normal to the face
and $\lambda$ is the bending rigidity. The sums extend to nearest
neighbours.  Eq.~\ref{action} describes {\it phantom} surfaces since
it does not include self-avoidance. The surface is a rhombus with free
boundaries cut out of a triangular lattice.
In the case of a spring of length $a$,
($l_{\rm x x'}$) has to be replaced by ($l_{\rm x x'} - a$). In this case
our model would closely resemble the one of \cite{KN} discussed
above.

We focused our analysis on the following observables: the {\it
specific heat},
\begin{equation}
C_v = \frac{\lambda^2}{V} \left(\langle S_e^2 \rangle - \langle S_e
\rangle^2 \right).
\label{specheat}
\end{equation}
Here $S_e$ is the bending energy term considered above and $V$ is the
total number of vertices.

The {\it radius of gyration},
\begin{equation}
R_g^2 = \frac{1}{3V} \langle \sum_i {\bf r}_i' \cdot {\bf r}_i' \rangle.
\label{gyration}
\end{equation}
Here ${\bf r}_i'$ is the position of the node $i$ in the embedding
space referred to the center of mass.  This observable measures the
physical extent of the surface and its scaling behavior with system
size defines the size (Flory) exponent $\nu$, via the relation $R_g
\propto L^{\nu}$. The exponent $\nu$ is related to the Hausdorff
dimension $d_H$ via the relation $\nu = 2/d_H$.

The {\it eigenvalues of the inertia tensor}; these eigenvalues give
information on the shape of the surface and how it scales with system
size.  They are obtained by diagonalizing the anisotropic part of the
inertia tensor
\begin{equation}
A_{ab} = \sum_{i} {r}_a'(i) {r}_b'(i)
\label{anisotensor}
\end{equation}
where $a,b$ refer to the components of the vector ${\bf r'}$.
        
The {\it normal-normal correlation function},
\begin{equation}
G(R) = \langle \frac{1}{N} \sum_{\vert\alpha\vert = R} \left(
\vec{n}_{\alpha} \cdot \vec{n}_O\right) \rangle.
\label{corrfunction}
\end{equation}
Here the sum extends to all triangles of the surface which have a
geodesic distance $R$ from the center of the surface $O$.  The angle
brackets represent the Monte-Carlo average.

\begin{figure}[t]
\epsfxsize= \textwidth \epsfbox{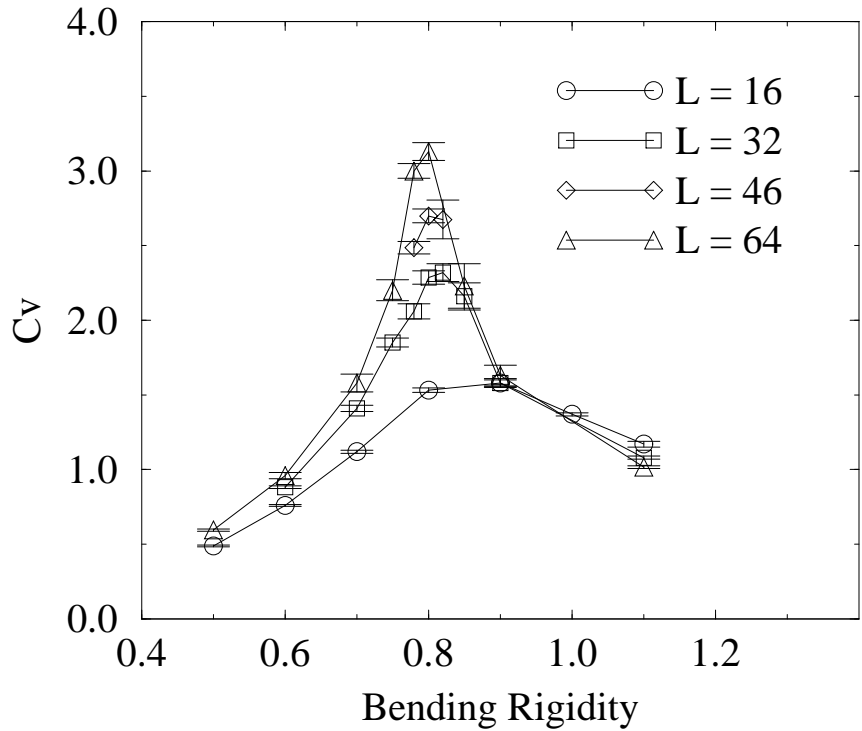}
\protect\caption{\protect\label{SPECFIG}The specific heat as a
function of the bending rigidity.}
\end{figure}

\begin{figure}[t]
\epsfxsize= \textwidth \epsfbox{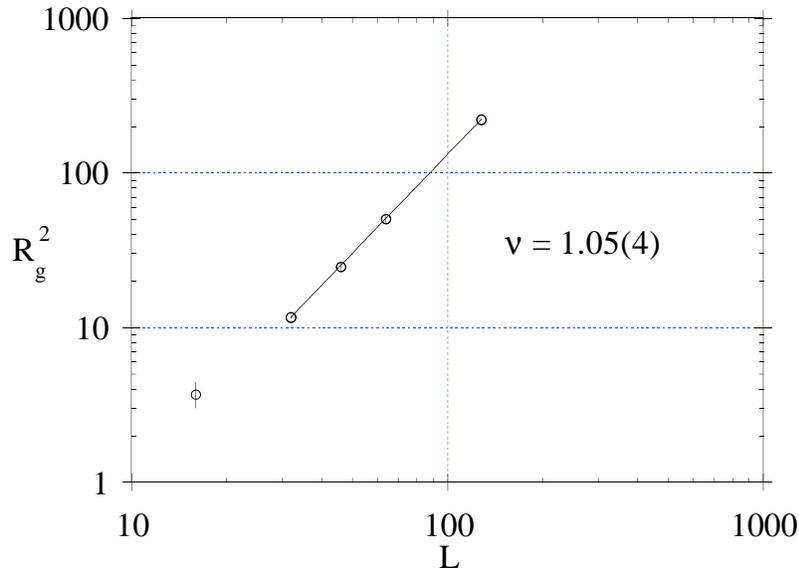}
\protect\caption{\protect\label{RADFIG}The scaling of the radius of
gyration squared with system size in the cold phase($\lambda = 1.1$).
The fit gives a value of $\nu = 1.05(4)$.}
\end{figure}

\section{Theoretical Predictions}

A self-consistent perturbation theory analysis of the continuum model
\cite{LDR} yields predictions for the critical exponents. The
exponents of interest are the size (Flory) exponent $\nu$ and the
roughness exponent $\xi$. The roughness exponent is defined by the
scaling of the minimum eigenvalue of the inertia tensor
(\ref{anisotensor}), 

\begin{equation}
\label{eicscaling}
\lambda \propto L^{2\xi}.
\end{equation}

At the critical point the theory predicts $\nu~=~\xi~=~0.732$ while in
the cold phase $\nu~=~1$ and $\xi~=~0.59$. 

As far as the normal-normal correlation is concerned the only
analytical result is for $\lambda = 0$ (gaussian model).  In this case
the correlation function follows a decay law $G(R) \propto - 1/R^4$.

\section{Numerics and Results}
We performed Monte-Carlo simulations of systems of sizes $16^2$ to
$128^2$ vertices. We used the single hit Metropolis algorithm.  The
largest lattice was simulated on a MASPAR MP1 massively
parallel processor, while all other sizes were simulated on
workstations.  We gathered statistics of the order of 30--50$\times
10^6$ sweeps per data point for the largest lattices (64 and 128).
Our statistics are comparable for the smaller lattices.

As can be seen in Fig.~\ref{SPECFIG} the specific heat $C_v$ shows a
growing peak with system size.  Presently our statistics are not yet
sufficient to allow a reliable estimate of the exponent $\alpha$ which
characterizes the growth. Preliminary fits indicate a value of $\alpha
\approx 0.5$ consistent with the value obtained in \cite{HW,WS} using
the same method.  Our best estimate for the critical value of the
coupling is around $\lambda \simeq 0.79$. Work is currently under way
to gather better statistics and perform a Ferrenberg-Swendsen type
analysis.

Fig.~\ref{RADFIG} shows the radius of gyration versus system size at a
fixed value of $\lambda$ (1.1).  The data fits well to a scaling
ansatz with $\nu = 1.05(4)$, as expected in the flat phase.  In the
crumpled phase the data does not fit a power law behavior,
indicating, as expected \cite{KN,KKN}, a log-like scaling ($d_H =
\infty$). Our estimate of the critical coupling is not precise enough
to allow for a fit to $R_g$ at the transition.

\begin{figure}[t]
\epsfxsize= \textwidth \epsfbox{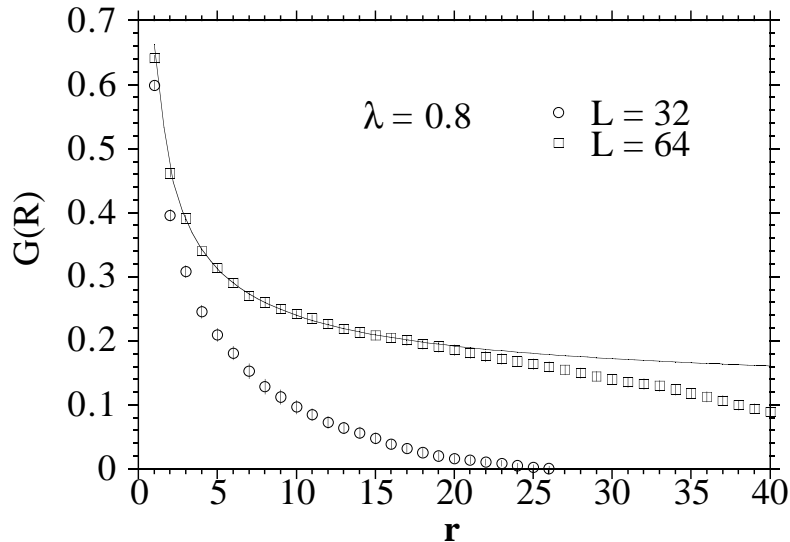}
\protect\caption{\protect\label{CORR1FIG}The normal-normal correlation
function at $\lambda = 0.8$ (around the phase transition).}
\end{figure}

\begin{figure}[t]
\epsfxsize= \textwidth \epsfbox{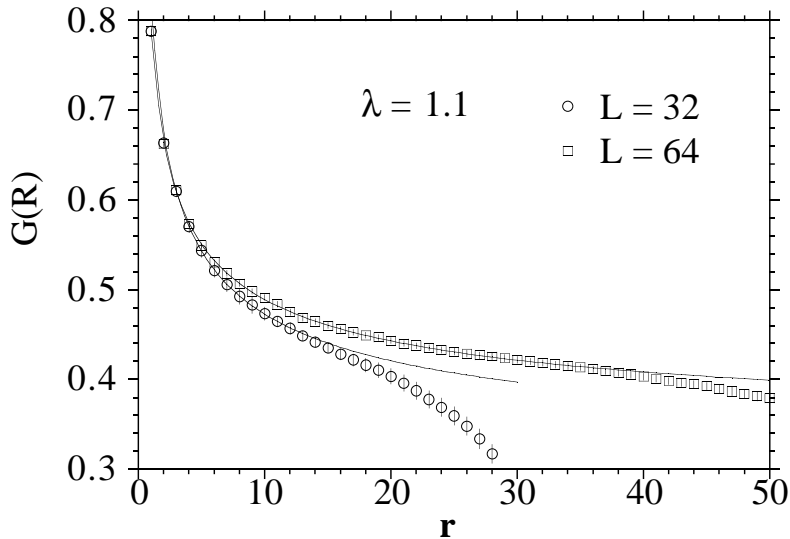}
\protect\caption{\protect\label{CORR2FIG}The normal-normal correlation
function at $\lambda = 1.1$ (in the cold phase).}
\end{figure}

Figures \ref{CORR1FIG} and \ref{CORR2FIG} show the normal-normal
correlation around and above the phase transition respectively.
Fig.~\ref{CORR1FIG} demonstrates the effect of finite-size corrections
to the value of the critical coupling.  For the smaller volume $L =
32$ the correlation decays to zero with $r$, but fits indicate a
non-zero asymptotic value for $L = 64$.  One possible reason is that,
due to the volume dependence of the pseudo-critical coupling, the
smaller volume is in the crumpled phase while the larger is in the
cold (flat) phase.  Note that for large $r$ the correlations
will always decay to zero because of our choice of boundary
conditions.  This is a finite size effect and the data close to the
boundary has to be excluded from the fits.

Figure \ref{CORR2FIG} shows the correlation in the cold phase.  The
data for $L = 64$ fits well to a behavior
\begin{equation}
G(R) \simeq \frac{1}{R^{a}} + b
\label{corrfit}
\end{equation}
with $a = 0.51(1)$ and $b = 0.3(1)$. Data for higher values of
$\lambda$ and of the system size $L$ show a consistent behavior.

This is very important---the presence of a non-zero asymptote for
the normal-normal correlation function indicates that the normals
remain ordered on a macroscopic scale.

This result supports the existence of a stable flat phase.  Our
present focus is on the precise nature of this phase; in particular
we would like to know if there is a well defined roughness exponent
$\xi$ and if so how it depends on the bending rigidity.  The
existence of a flat phase with a roughness exponent independent of
$\lambda$ would indicate that the entire flat phase $\lambda >
\lambda_c$ is critical and that this model is in the same universality
class as \cite{AL,GDLP,AGL}.

Another possibility is that the roughness exponent will depend on
$\lambda$.  In this case the flat phase would still be stable but not
critical. 

A possible interpretation of the stability of the flat phase is that
the model discussed here dynamically generates a non-zero length scale
which can serve to define non-zero renormalized elastic constants. 
This simple model could then be used to study the properties of
physical (polymerized) membranes in regimes in which self-avoidance is
irrelevant. 

We are grateful for the use of  NPAC computational facilities. 
The research of M.B.~and M.F.~was supported by the Department of
Energy U.S.A.~under contract No.~DE-FG02-85ER40237.  M.B.~would also
like to acknowledge support under NSF grant No.~PHY89-04035 from
the ITP at Santa Barbara, where some of this work was carried out. 
S.C.~and G.T.~were
supported by research funds from Syracuse University. The
research of K.A.~was supported by the Department of Energy U.S.A.~ under
contract DE-FG05-86ER-40272.
K.A.~acknowledges IFT at Gainesville where part of this work was carried out.

\end{document}